\documentclass{PoS}

\usepackage{array}
 \usepackage{multicol}

\title{Parton energy loss effect on Z+jet production in high-energy nuclear collisions}

\ShortTitle{Z+jet correlations in heavy-ion collisions}

\author{\speaker{Shan-Liang Zhang} $^a$, Tan Luo $^a$, Xin-Nian Wang$^{ab}$, Ben-Wei Zhang$^a$  \\
        \llap{$^a$} Institute of Particle Physics and Key Laboratory of Quarks and Lepton Physics (MOE), Central China Normal University, Wuhan 430079, China\\
        \llap{$^b$} Nuclear Science Division Mailstop 70R0319, Lawrence Berkeley National Laboratory, Berkeley, CA 94740\\
        E-mail: \email{zhangshanl@mails.ccnu.edu.cn},  \email{luotan@mails.ccnu.edu.cn},  \email{xnwang@lbl.gov}, \email{bwzhang@mail.ccnu.edu.cn}}

\abstract{We give a report of medium modification of Z+jet correlations in Pb+Pb collisions at the Large Hadron Collider using Sherpa to generate initial Z+jet at next-leading-order matrix element matched parton shower, and the Linear Boltzmann Transport Model for jet propagation in the expanding quark-gluon-plasma. Our
numerical calculations show excellent agreement with all available observables of Z+jet simultaneously in both proton + proton and Pb+Pb collisions. Our results can well explain the shift of momentum asymmetry $x_{jZ}=p_T^{jet}/p_T^Z$ as well as its mean values, the suppression of the jet yields per Z trigger $R_{jZ}$ and the modification of azimuthal angle correlation $\Delta \phi_{jZ}$.  We also demonstrate that it is the energy loss effect on multi-jets from high-order corrections that leads to the suppression of the Z+jet correlation at small azimuthal angle difference $\Delta \phi_{jZ}$ and at small $x_{jZ}$. The jet shape reflecting transverse momentum distribution inside the jet is also calculated, which indicates that large fraction of jet energy is carried away from the jet axis in Pb+Pb collisions.}

\FullConference{International Conference on Hard and Electromagnetic Probes of High-Energy Nuclear Collisions\\
		30 September - 5 October 2018\\
		Aix-Les-Bains, Savoie, France}

\begin{document}

\section{introduction}
 Jet production in association with Z boson provides an ideal probe of the properties of the quark-gluon plasma (QGP)~\cite{Neufeld:2010fj}.
 The outgoing partons interact strongly with the hot/dense medium and lose energy in the QGP~\cite{Gyulassy:2003mc}, while Z boson will not participate in the strong-interactions directly, escaping the QGP unscathed. Besides, Z boson is free from fragmentation and decay due to its large mass ($M_Z=91.18$ GeV). Therefore, the Z boson transverse momentum closely reflects the initial energy of the associated parton that fragments into the final-state jet.

Z+jet correlations on transverse momentum asymmetry $x_{jZ}=p_T^{jet}/p_T^Z$ as well as its mean value $\langle x_{jZ}\rangle$, jet yields per Z trigger $R_{jZ}=N_{jZ}/N_Z$, and azimuthal correlation $\Delta \phi_{jZ}=|\phi_{jet}-\phi_Z |$ both in proton+proton (p+p) and lead+lead (Pb+Pb) collisions at 5.02 TeV have been measured by CMS  experiment~\cite{Sirunyan:2017jic}. It is noted when computing  $\Delta \phi_{jZ}$, the next-leading-order (NLO) calculations suffer divergence in the region $\Delta \phi_{jZ}\sim \pi$, because of soft/collinear radiation. Furthermore, even though leading-order (LO) matched parton shower (PS) calculations have already contained some high-order corrections from  real and virtual contributions, it is short of additional hard radiation from high-order matrix element calculations, as a consequence of which, it underestimates the azimuthal angle correlation at small angle difference region~\cite{Zhang:2018urd,Dai:2018mhw}.  Motivated by this, we present in this talk a state-of-art calculations of Z+jet~\cite{Zhang:2018urd}, with p+p baseline compuated at NLO+PS with Sherpa~\cite{Gleisberg:2008ta}, and the Linear Boltzmann Transport (LBT) model~\cite{He:2015pra} for jet propagation in heavy-ion collisions.

\section{Model setup for Z+jet in heavy-ion collisions}

 Initial reference Z+jet events in p+p collisions is simulated at NLO matrix element perturbative calculations matched to the resummation of parton shower~\cite{Hoche:2010kg, Hoeche:2012yf} within a Monte Corlo event generator Sherpa~\cite{Gleisberg:2008ta} at $\sqrt {s_{NN}}=5.02 $ TeV. NLO +PS calculations of azimuthal angle correlation and momentum asymmetry  for Z+jet agree well with experiment data~\cite{Sirunyan:2017jic} in all kinetic ranges in p+p collisions~\cite{Zhang:2018urd}. EPPS16 modified npdfs is used to study cold nuclear matter effects, but no modifications is observed as in~\cite{Ru:2016wfx}.



The Linear Boltzmann Transport (LBT) model is then used to simulate the propagation, energy attenuation of, and medium
response induced by jet partons in the quark-gluon plasma~\cite{He:2015pra}. LBT is based on a Boltzmann equation ~\cite{He:2015pra}:
\begin{equation}
p_a\cdot\partial f_a(p_a)=-\frac{1}{2}\int\sum _{i=b,c,d}\frac{d^3p_i}{(2\pi)^32E_i}\times[f_af_b-f_cf_d]|M_{ab\rightarrow cd}|^2\times S_2(s,t,u)(2\pi)^4\delta^4(p_a+p_b-p_c-p_d)
\end{equation}
where $f_i$ are phase-space distributions of partons, $S_2(s,t,u)$ is Lorentz-invariant regulation condition. Elastic scattering is introduced by the complete  set of $2 \rightarrow 2$ matrix element $|M_{ab\rightarrow cd}|$, and the inelastic scattering is described by high-twist formalism for induced gluon radiation ~\cite{Guo:2000nz, Zhang:2003wk, Schafer:2007xh}.

 \section{Numerical results}

 To compare with the experimental data, we select the Z boson and jets according to the kinematic cut adopted by CMS~\cite{Sirunyan:2017jic}. The information of the evolving bulk matter is provided by (3+1)D hydrodynamics~\cite{Pang:2012he}. The underlying event background energy is subtracted event-by-event for Pb+Pb collisions following the procedure applied in CMS~\cite{Kodolova:2007hd}, while no subtraction is applied in p+p collisions.

\begin{figure}
  \centering
   \includegraphics[scale=0.23]{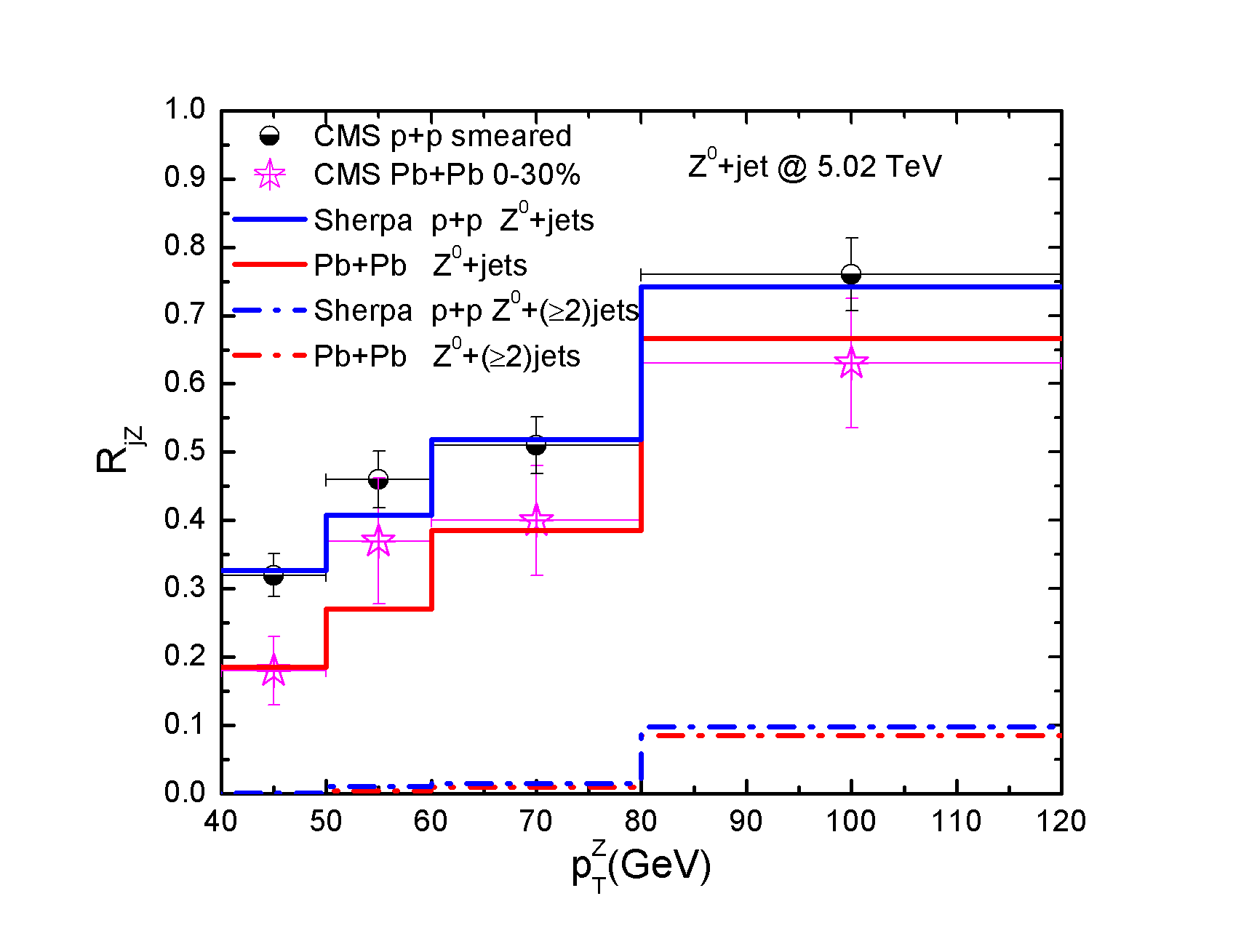}
   \includegraphics[scale=0.23]{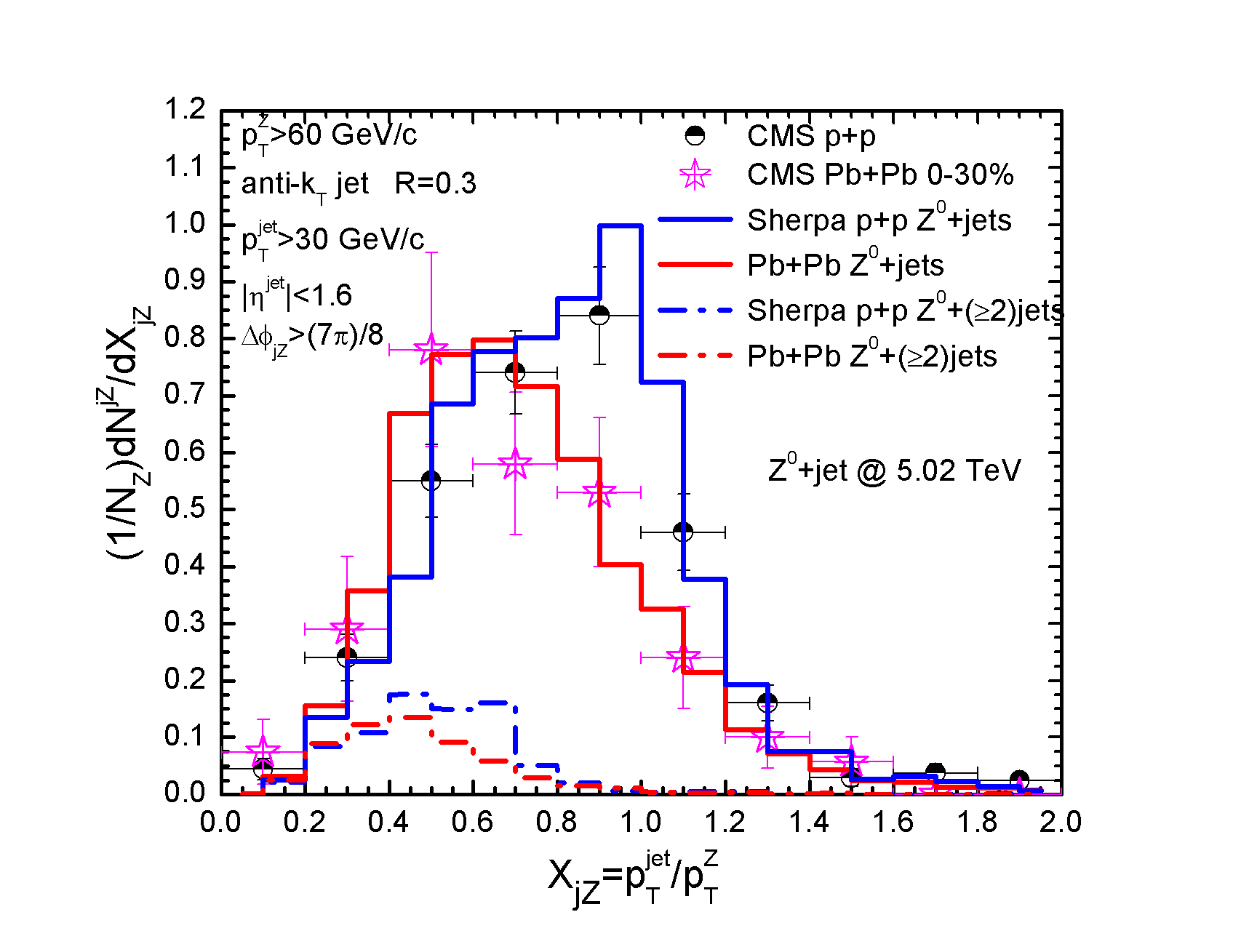}
         \vspace{-10pt}
  \caption{(Color online) (Left) Distributions of  $R_{jZ}=N_{jZ}/N_Z$ (left) and $x_{jZ}=p_T^{jet}/p_T^Z$ (right)
  in central Pb+Pb collisions and p+p collisions at $\sqrt {s_{NN}}=5.02 $ TeV.
  }\label{rjz}
\end{figure}

 We first fix the only parameter $\alpha_s$ that controls the strength of jet-medium interactions via the comparison with the CMS data of Z+jets~\cite{Sirunyan:2017jic}.  When $\alpha_s$ is set to 0.2, our numerical results of average number of jet partners per Z boson $R_{jZ}$ in central Pb+Pb collisions show well agreement with CMS data as in Fig.~\ref{rjz} (left). $R_{jZ}$ is overall suppressed in Pb+Pb, because a large fraction of jets lose energy and then shift their final transverse momenta below the threshold $p^{jet}_T=$ 30 GeV.

The imbalance in the transverse momentum of the associated jet relative to that of recoiled Z boson  $x_{jZ}=p_T^{jet}/p_T^Z$  is presented in Fig.~\ref{rjz} (right).  Compared to p+p collisions,  there is a significant displacements of the peak value of $x_{jZ}$ towards a smaller value in Pb+Pb, due to jet energy loss in the medium while the transverse momentum of Z boson is unattenuated. Multi-jets processes are rather important when $x_{jZ}<0.5$,  where the multi-jets energy can hardly exceed half of the energy of Z boson in the phase space $\Delta \phi_{jZ}\ge \frac{7\pi}{8}$.

\begin{figure}
  \centering
   \includegraphics[scale=0.23]{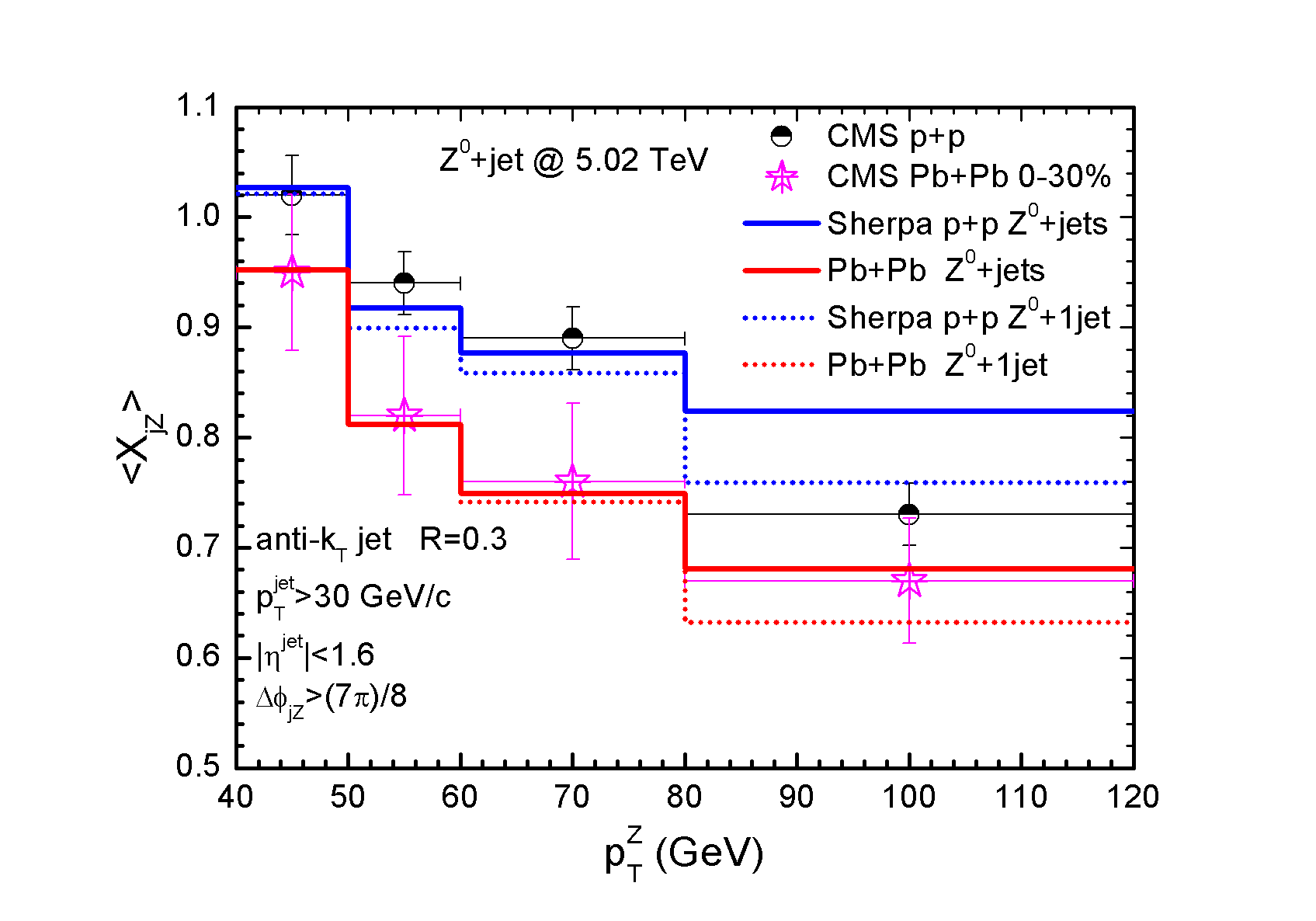}
   \includegraphics[scale=0.23]{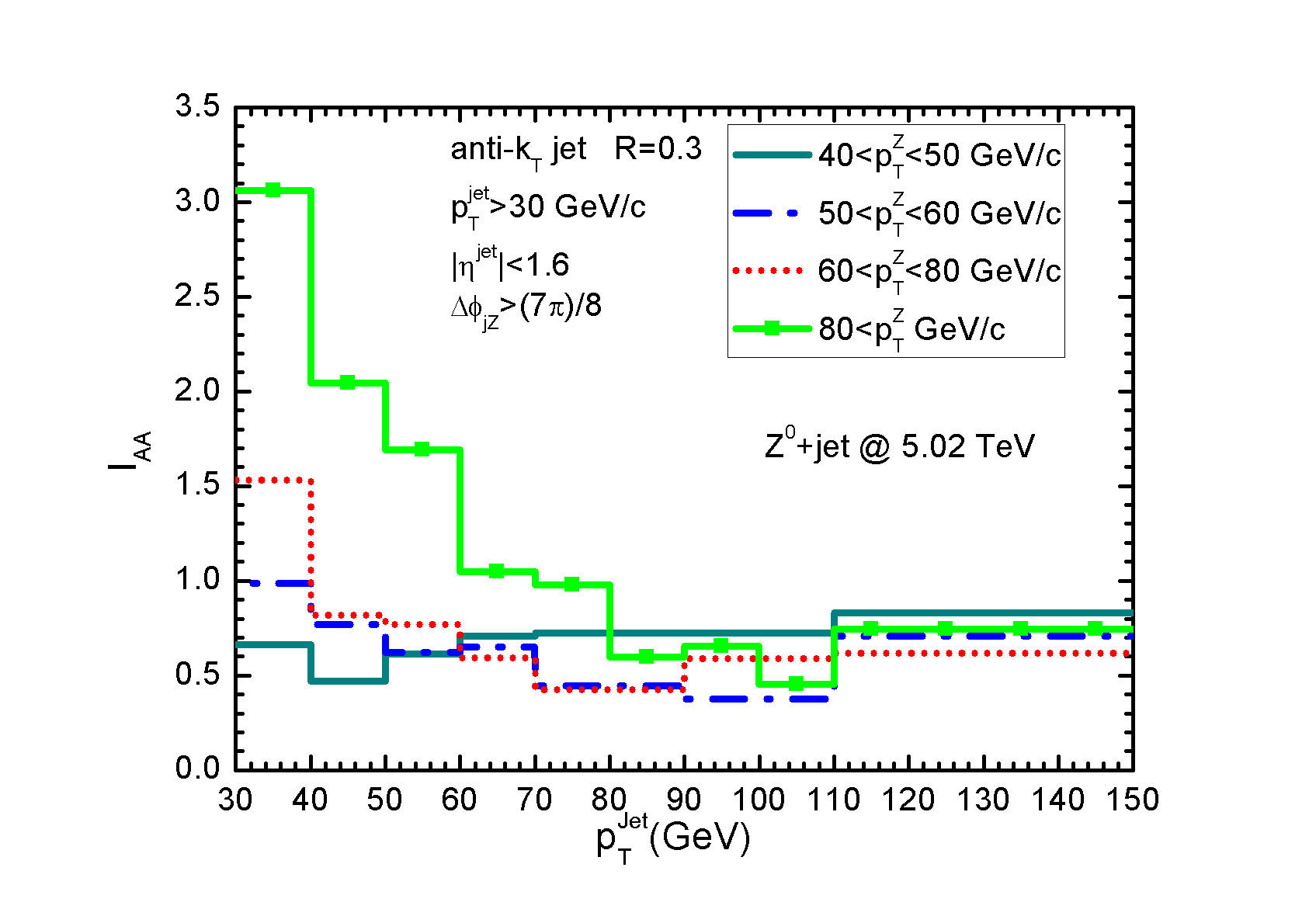}
      \vspace{-10pt}
  \caption{(Color online) (Left) Mean value of momentum imbalance $\langle x_{jZ}\rangle$ of Z+jet in central Pb+Pb (red) and p+p collisions (blue) at $\sqrt{s_{NN}}=5.02$ TeV.
   (Right) $I_{AA}$ as a function of $p_T^{jet}$ in different $p_T^Z$ bins.  }\label{xjz}
\end{figure}

To quantify the relative shift between p+p and 0-30$\%$ central Pb+Pb collisions, the mean value of the momentum asymmetry $\langle x_{jZ}\rangle$ is calculated and shown in Fig.~\ref{xjz} (left). It is much smaller in Pb+Pb relative to p+p collisions. Fig.~\ref{xjz} (right) plots the nuclear modification factor $I_{AA}=(dN^{Pb+Pb}/dp_T^{jet})/(dN^{p+p}/dp_T^{jet})$ of the leading jet tagged by Z boson.  An enhancement is observed at $p_T^{jet} < p^Z_T$ region,  and  a suppression in  $p_T^{jet} > p^Z_T$ region. We find $I_{AA}$ is quite sensitive to the kinematic cut due to the steep falling cross section in the kinematic cut window  .

\begin{figure}[tpb]
 \begin{center} \centering

   \includegraphics[scale=0.21]{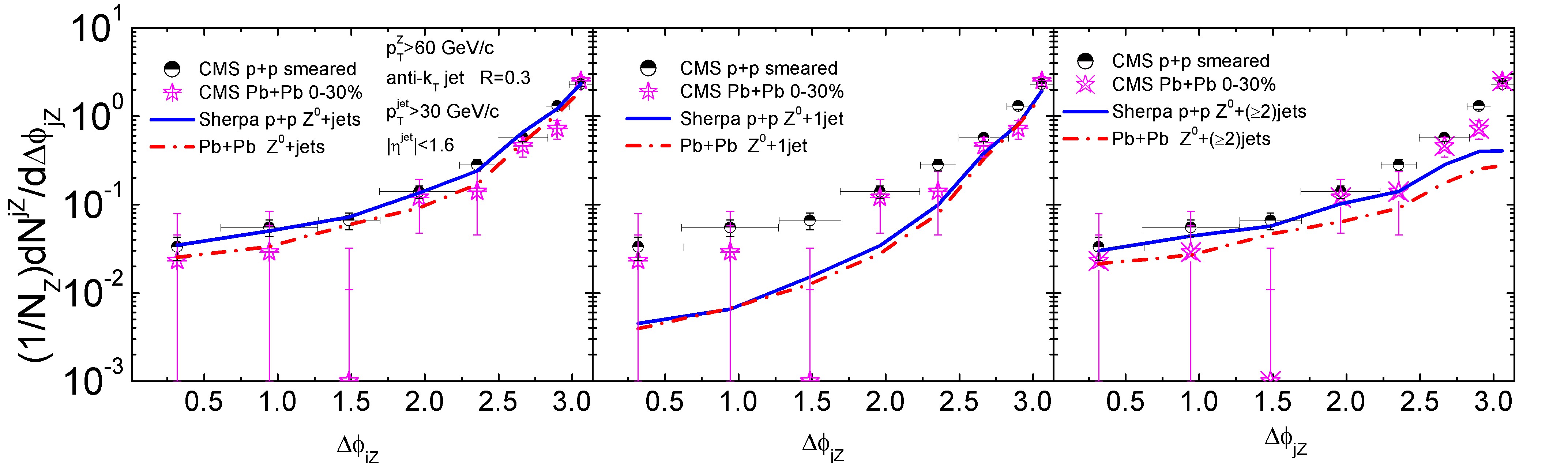}
 \end{center}
   \vspace{-10pt}
  \caption{(Color online) Z+jet azimuthal angle correlation $\Delta\phi_{jZ}=|\phi_{jet}-\phi_Z|$ (left), and the contributions  from Z plus only one jet (middle), and Z plus more than one jets (right) both in central Pb+Pb collisions and p+p collisions at $\sqrt{s_{NN}}=5.02$ TeV.   }\label{phi}
\end{figure}

Z+jet azimuthal angle correlation $\Delta\phi_{jZ}=|\phi_{jet}-\phi_Z|$  in p+p and Pb+Pb are shown in Fig.~\ref{phi} (left).  It is moderately suppressed in Pb+Pb collisions. To illustrate the suppression mechanism, separated contributions from Z+1jet and Z associated with more than one jets in both p+p and Pb+Pb collisions are revealed in Fig.~\ref{phi}. We see Z + 1jet dominates in large angle region and there is no significant difference between p+p and Pb+Pb collisions. These processes mainly come from  the LO ME  and the  azimuthal angle decorrelation from which is dominated by soft/collinear radiation. The transverse momentum broadening of jets due to jet-medium interaction is negligible at such high energy scale. The right panel of Fig.~\ref{phi} illustrates  that  Z+ multi-jets processes are considerably suppressed in Pb+Pb collisions.

\begin{figure}
  \centering
  \includegraphics[scale=0.085]{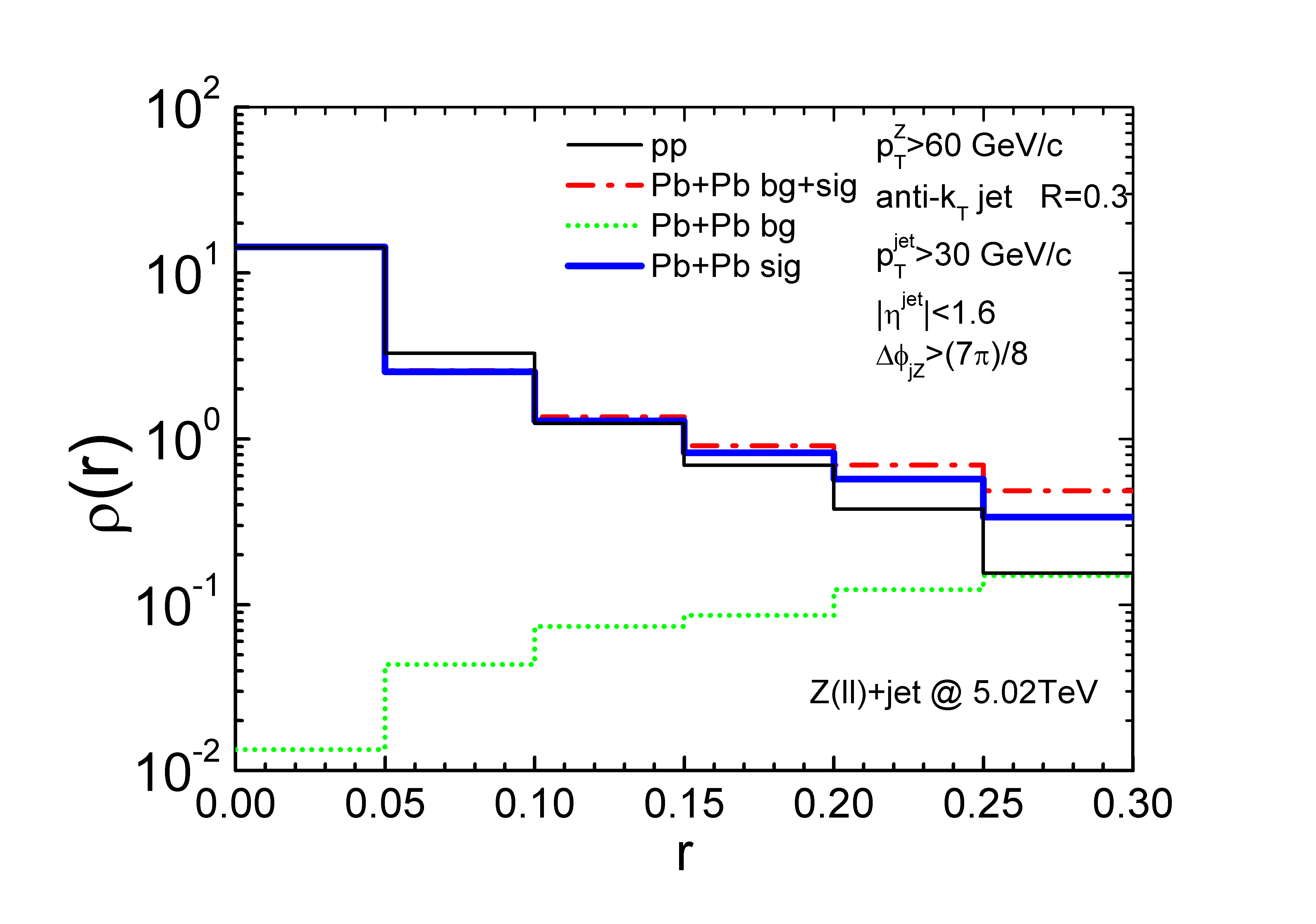}
  \includegraphics[scale=0.085]{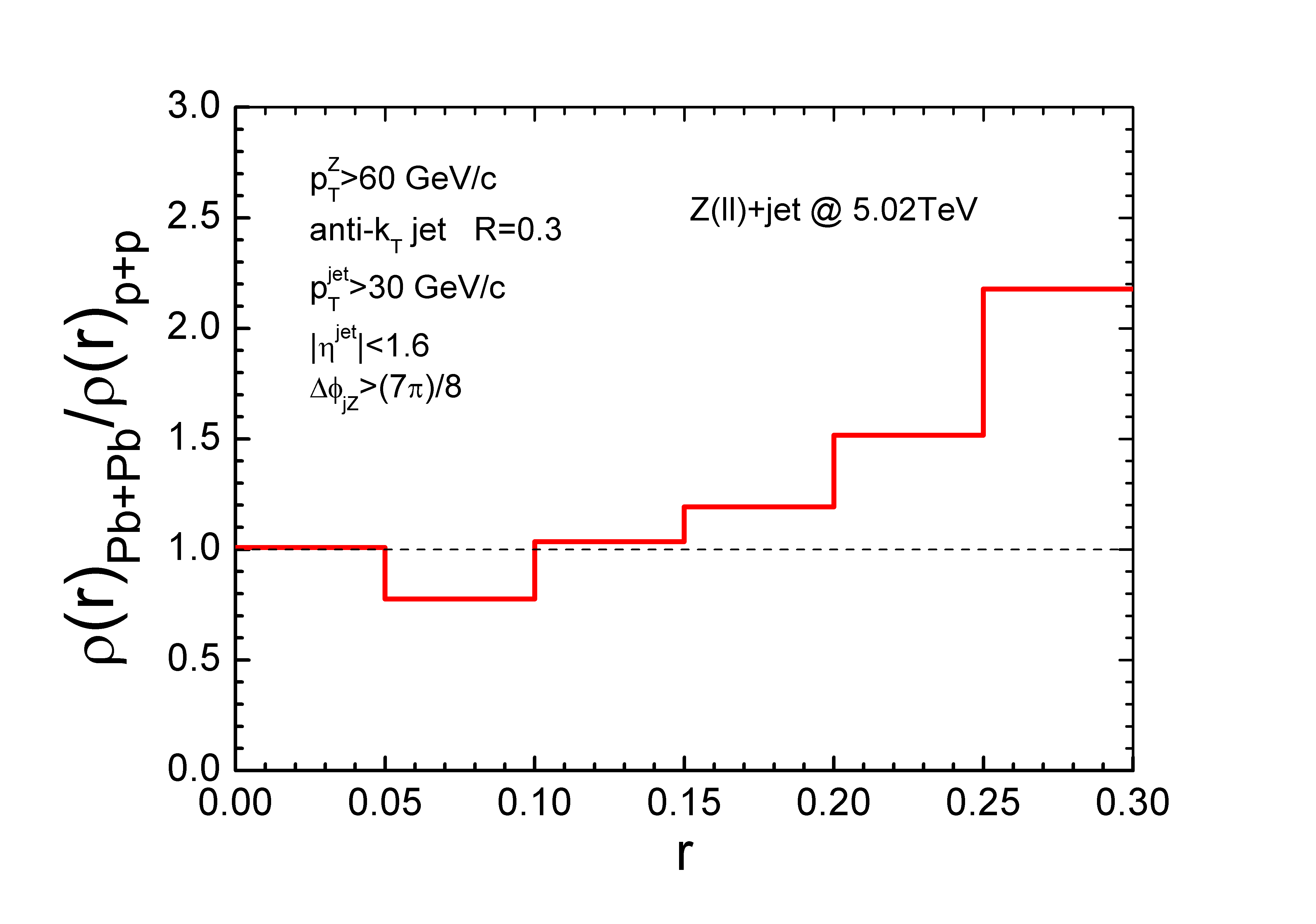}
  \caption{ (Color online) Differential jet shape $\rho (r)$ of jets triggered by Z boson in 0-30$\%$ central Pb+Pb and p+p collisions at $\sqrt s_{NN}=5.02$ TeV  as well as the ratio of jet shape in central Pb+Pb to that in p+p collisions.}\label{jetshape}
\end{figure}

In addition to Z+jet correlations, we calculated the differential jet profile which describes the radial distribution of transverse momentum inside the jet
 cone~\cite{Vitev:2008rz}.
 The differential jet shape in Pb+Pb and p+p collisions are displayed in Fig.~\ref{jetshape}. The result is normalized to unity over $r<0.3$. We see that, a large fraction of jet energy is carried in the core of the jet within $r<0.1$. To quantify the modification, we present the ratio of the jet shape in Pb+Pb to that in pp collisions in Fig.~\ref{jetshape} (right). We observe a deletion in the region $ 0.05<r<0.1$ and a enhancement at large radius $r>0.1$. It indicates that the energy is redistributed in Pb+Pb collisions due to jet-medium interactions and large amount of jet energy is carried by particles far away from the jet axis.

This work has been supported by NSFC of China with Project Nos. 11435004, and NSF
 under grant No. ACI-1550228 and U.S. DOE under Contract No. DE-AC02-05CH11231.

\end{document}